\documentclass[useAMS,usenatbib,letterpaper]{mn2e}
\usepackage{times}
\usepackage{amssymb}
\usepackage{lscape,graphicx}

\title[Clustering of passive galaxies]{The evolution of galaxy
  clustering since $z=3$ using the UKIDSS Ultra Deep Survey: the
  divergence of passive and star-forming galaxies}
\author[W. G. Hartley et al.]{W.~G. Hartley$^{1}$\thanks{E-mail:
    ppxwh1@nottingham.ac.uk}, O. Almaini$^{1}$, M. Cirasuolo$^{2}$,
  S. Foucaud$^{1,3}$, C. Simpson$^{4}$,
  \newauthor C.~J. Conselice$^{1}$, I. Smail$^{5}$,
  R.~J. McLure$^{2}$, J.~S. Dunlop$^{2}$, R.~W. Chuter$^{1}$,
  \newauthor S. Maddox$^{1}$, K.~P. Lane$^{6}$, E.~J. Bradshaw$^{1}$ \\
  $^{1}$School of Physics and Astronomy, University of Nottingham, University Park, Nottingham NG7 2RD \\
  $^{2}$Institute for Astronomy,
  University of Edinburgh, Royal Observatory, Edinburgh EH9 3HJ \\
  $^{3}$Department of Earth Sciences, National Taiwan Normal University, No. 88, Section 4, Tingzhou Road, Wenshan District, Taipei 11677, Taiwan \\
  $^{4}$Astrophysics Research Institute, Liverpool John Moores University, Twelve Quays House, Egerton Wharf, Birkenhead CH41 1LD  \\
  $^{5}$Institute for Computational Cosmology, Department of Physics, Durham University, Durham DH1 3LE\\
  $^{6}$Department of Astrophysics, University of Oxford, OX1 3RH\\
}

\begin{document}

\date{}

\pagerange{\pageref{firstpage}--\pageref{lastpage}} \pubyear{2010}

\maketitle

\label{firstpage}

\begin{abstract}

We use the UKIDSS Ultra-Deep Survey to trace the evolution of galaxy clustering to $z=3$. 
Using photometric redshifts derived from data covering
the wavelength range $0.3 - 4.5 {\rm \mu m}$ we examine this
clustering as a function of absolute K-band luminosity, colour and
star-formation rate. Comparing the deprojected clustering
amplitudes, we find that red galaxies are more strongly clustered than
blue galaxies out to at least $z=1.5$, irrespective of rest-frame K-band luminosity. 
We then construct passive and star-forming samples based
on stellar age, colour and star-formation histories calculated from
the best fitting templates. The clustering strength of star-forming
galaxies declines steadily from $r_0\simeq7h^{-1}$Mpc at $z\simeq2$ to
$r_0\simeq3h^{-1}$Mpc at $z\simeq0$, while passive galaxies have
clustering strengths up to a factor of two higher.
Within the passive and star-forming subsamples, however, we find very
little dependence of galaxy clustering on K-band luminosity. Galaxy
`passivity' appears to be the strongest indicator of clustering strength.
We compare these clustering measurements
with those predicted for dark matter halos and conclude that passive
galaxies typically reside in halos of mass $M\ge10^{13}M_{\odot}$ while luminous
star-forming galaxies occupy halos an order of magnitude
less massive over the range $0.5<z<1.5$. The decline in the clustering strength of
star-forming galaxies with decreasing redshift indicates a
decline in the hosting halo mass for galaxies of a given luminosity.
We find evidence for convergence of clustering in star-forming and
passive galaxies around $z\sim2$, which is consistent with this being
the epoch at which the red sequence of galaxies becomes distinct.
\end{abstract}


\section{Introduction}

The origin of the red sequence of galaxies, and more generally the bimodality
in the colour-absolute magnitude plane \citep{Visvanathan77}, have become increasingly
important questions in extragalactic astronomy. Since \cite{Dressler80} first
identified the trend for elliptical galaxies to be preferentially found in the
denser regions of clusters, evidence has grown to suggest that there are
crucial differences in the evolutionary histories of red and blue
galaxies at low redshift. Present day massive, passive galaxies are claimed to be largely in place by z=1 or
earlier \citep[e.g.][]{Yamada05,Cimatti06,Conselice07}, implying an extremely rapid period of star-formation. The
luminosity and mass functions of lower-mass and star-forming galaxies,
however, evolve significantly over the same time period
\citep{Bundy06,Cirasuolo10}. 

The early formation of stars in the most massive galaxies has become known as
downsizing \citep{Cowie96} and is in contrast to the well-established
hierarchical growth of the underlying dark matter distribution. These
massive, quiescent galaxies dominate present-day groups
and clusters and are thought to have formed at the
highest peaks of the dark matter density field. 
The expected clustering of the dark
matter is well understood (e.g. \citealt{Mo02}), and at any given
epoch the real-space clustering amplitude over mid to large scales is a monotonically
increasing function of dark matter halo mass. Measuring the clustering
properties for galaxy samples therefore provides a direct link from
the visible galaxies to the invisible dark matter in which they are
found. Furthermore, the growth of dark matter halos over time is also
well modelled, and so the evolution of the most massive systems
throughout cosmic history can be traced through the use of clustering
statistics.

Over the last few years the clustering technique has been applied to
complete samples of optically-selected galaxies, separated
by luminosity and colour.
\cite{Zehavi05} computed the angular auto-correlation functions of red and blue
volume-limited sub-samples of galaxies from the Sloan Digital Sky Survey
(SDSS; \citealt{York00}) for separations $<10h^{-1}$Mpc and redshifts, $z<1$. They found that the 
population of galaxies red in $g-r$ colour are more
strongly clustered than those with bluer colours ($r_0 = 5.7 {\rm ~and~}
3.6 h^{-1}$Mpc for red
and blue populations respectively). Moreover, they show that this difference
remains when the samples are further broken down into absolute R-band
magnitude sub-samples.

\cite{Meneux06} split the VIMOS-VLT Deep Survey (VVDS; \citealt{LeFevre05})
galaxies into red and blue spectral types by the method of \cite{Zucca06} and
computed their clustering out to $z \simeq 1.2$. They found that the
deprojected clustering scale length of the red spectral type galaxies
($r_0\sim5h^{-1}$Mpc) was
greater than that of the blue type galaxies ($r_0\sim3h^{-1}$Mpc)
at all redshifts investigated. \cite{Carlberg97} find similar results for
their sample at $z \simeq 0.6$. More recently \cite{Coil08} used the DEEP2
galaxy redshift survey to address the same question at $z=1$. They also found
that red-sequence galaxies are more strongly clustered
($r_0\sim 5 {\rm ~and~} 3h^{-1}$Mpc for red and blue samples), but additionally
showed that those galaxies with intermediate colours had an
intermediate clustering strength. They computed the relative bias between
their sub-samples, finding a smoothly increasing bias with rest-frame (U-B)
colour up to the red sequence.

Recently, two further deep surveys have been used to study the
clustering of galaxies in the range $0.2<z<1.2$: the
Canada-France-Hawaii Telescope Legacy Survey \citep[CFHTLS,
][]{Ilbert06} and the zCOSMOS field
\citep{Lilly07}. \cite{McCracken08} used the CFHTLS, splitting the
galaxy sample by template-fit spectral type and B-band
luminosity. Based on photometric redshifts, they find that the
difference in clustering strength between galaxies fit by early type
and late type templates is roughly constant across all redshifts
studied. The luminosity dependence at a given redshift, however, is
minimal. Intriguingly, the early type population at $z\sim0.5$ shows
hints of an inverse dependence of luminosity on clustering
strength. The authors attribute this behaviour to fainter red galaxies
being preferentially located in cluster environments. \cite{Meneux09}
meanwhile, use the spectroscopic zCOSMOS data, finding a
weak dependence on luminosity which remains when stellar mass is used
instead.

Each of these studies has been based upon an optical colour selection,
which limits these studies to $z\le1$. Beyond this
redshift optical studies become increasingly biased against passive
and dusty galaxies and so are only able to probe the clustering of
unobscured star-forming galaxies. The relative clustering strengths for
complete samples of passive and star-forming galaxies are therefore
poorly understood beyond $z\sim1$. Infrared survey data are not biased
in this way and in addition, at $z<2$, the K-band luminosity is a reasonable
proxy for the stellar mass of a galaxy. Infrared-selected samples are
therefore vital in the study of high redshift galaxy clustering.

Utilising infrared data,
the high redshift progenitors of nearby high-mass, passive systems have been suggested as being 
passively-selected {\it BzK} galaxies \citep{Daddi04,Kong06,Hartley08}, 
ultra-compact z=2 red galaxies \citep[e.g.][]{Zirm07,Toft07,Cimatti08}, sub-mm galaxies
\citep{Lilly99,Swinbank06}, $24\mu$m sources selected at $z>1$ \citep[e.g.][]{Magliocchetti08}, distant red galaxies
(DRGs) \citep{Franx03,Conselice07,Foucaud07}, extremely red objects (EROs)
\citep{Roche02,McCarthy04} and Ultra-luminous infrared galaxies
(ULIRGS) \citep[e.g.][]{Farrah06}. Each
of these populations have been found to cluster very strongly, with
$r_0$ values in the range $7<r_0<14 h^{-1}{\rm Mpc}$
\citep{Roche02,Blain04,Farrah06,Foucaud07,Magliocchetti08,Weiss09}.

In \cite{Hartley08} (henceforth H08) we used the colour selection criteria of
\cite{Daddi04} to investigate the clustering properties of passive and
star-forming galaxies at $1.4 < z < 2.5$. We showed that, when limiting the
samples to $K_{AB}<23$, the passive galaxies were significantly more
clustered than star-forming galaxies. This result was found in
angular clustering and also when deprojected to find the real-space clustering
strength; with clustering scale lengths of $\sim15 h^{-1}$Mpc for the
passive sample (pBzK) and $\sim7 h^{-1}$Mpc for the star-forming
sample (sBzK).

Although all of these different galaxy populations overlap to some extent \citep{Lane07},
their biases differ and they are sensitive to different epochs. It is
therefore difficult to compare these samples with low redshift samples or
to study global galaxy evolution.
The analysis in H08 in particular left some open questions. We established that passive
galaxies are more clustered even at such high redshifts, but the
passive population is exclusively bright and the star-forming galaxies
exhibited a strong limiting-magnitude dependence on clustering
strength. It is therefore not clear whether
it is luminosity or passivity that is more significantly correlated with clustering
strength. Even at low-redshift debate remains over such a distinction
\citep{Norberg02}, complicated by the dominance of passive systems at
the bright end of the luminosity function.

\cite{Williams09} used a two-colour rest-frame selection to select a
quiescent cloud and star-forming track over the range
$1<z<2$. Using this novel selection, they compute the 
clustering scale lengths of their two samples, finding very
similar lengths to those of the BzK selections in H08. They also
report evidence for a mild evolution in clustering with luminosity for
the star-forming galaxies by splitting their samples into two
luminosity bins. \cite{Coil06} also find, at $z=1$, only a modest dependence for 
clustering strength to increase with
B-band luminosity, covering $0.7L^* < L < 1.4L^*$.

A remaining open question is to determine when the passive population is built up.
Though we found in H08 that this build-up occurs over $1.4 < z < 2.5$,
and \cite{Williams09} find similar evidence for their sample at $1 < z
< 2$, these
ranges in redshift are very broad. Can we identify a narrow epoch at which the clustering
strengths of passive and star-forming galaxies of a given luminosity are
equal? At this epoch passive and star-forming galaxies would occupy
the same mass of dark matter halos and differences in their internal processes
could be clearer. In this paper we attempt to disentangle the
effects on clustering strength due to passivity and luminosity and to further
refine the redshift range over which the passive population of galaxies is first established.

Where relevant, we adopt a concordance cosmology in our analysis; $\Omega_M =
0.3$, $\Omega_{\Lambda} = 0.7$, h$ = {\rm H}_0/100 = 0.7 ~{\rm
  kms}^{-1}{\rm Mpc}^{-1}$ and $\sigma_8 =
0.9$. All magnitudes are given in the Vega system unless otherwise stated. The
remainder of the paper is organised in the following way: in the
following section we provide details of the data sets used, derived
quantities and method of selecting passive galaxies. In section 3 we
compute the clustering strengths of these samples and then discuss
their implications in section 4. We conclude in section 5.

\section[]{Data sets, derived quantities and sample definitions}

The analysis we present in this work is based primarily on the third
data release (DR3) of the ongoing UKIRT (United Kingdom Infra-Red Telescope)
Infrared Deep Sky Survey, Ultra-Deep Survey (UKIDSS UDS,
\citealt{Lawrence07}). UKIDSS comprises five complementary sub-surveys
with the UDS being the deepest, targeting a depth of $K=23$
over a single, 4-pointing mosaic of the Wide-Field Camera (WFCAM,
\citealt{Casali07}): an area of $0.88\times0.88$ degrees. UKIDSS began 
in the spring of 2005 and the UDS has seen several
data releases: an early release of the first few weeks of data (EDR,
\citealt{Dye06}), a one-year release (DR1, \citealt{Warren07}), which
we used in \cite{Hartley08}, and the more recent three-year data release
(DR3). 

The principle improvement from the DR1 to the DR3 is the addition of
H-band data. Additional data in the K-band also provides an incremental
improvement in depth. The $5\sigma$ depths within
$2\arcsec$ diameter apertures for the DR3 in J, H and K-bands are $22.8$ (AB $23.7$),
$22.1$ (AB $23.5$) and $21.8$ (AB $23.7$) respectively. These depths make it the deepest
single-field near infrared survey of its area to date. For details of
the stacking procedure, mosaicing, catalogue extraction and depth
estimation we refer the reader to Almaini et al. (in prep.) and
\cite{Foucaud07}. 

In addition to near infrared data, the field is covered by extremely
deep optical data in the B, V, R, i$^{\prime}$ and z$^{\prime}$-bands
from the Subaru-XMM Deep Survey (SXDS,
\citealt{Sekiguchi05,Furusawa08}), {\it Spitzer} data reaching $5\sigma$
depths of 24.2 and 24.0 (AB) at $3.6\mu$m and $4.5\mu$m
respectively from a recent {\it Spitzer} Legacy Program (SpUDS, PI:Dunlop) and
U-band data taken with CFHT Megacam. The SXDS 
utilised Suprime-cam on the Subaru telescope, achieving depths of 
$B_{AB} = 28.4, ~V_{AB} = 27.8, ~R_{AB} = 27.7, ~i_{AB}^{\prime} =
27.7$ and $ z_{AB}^{\prime} = 26.7$ \citep[$3\sigma$,
  $2\arcsec$]{Furusawa08}. Border regions, areas around bright stars
and obvious cross-talk artifacts were masked and any sources within
such regions were discarded. The co-incident area with the UDS
taking masking into account is $0.63$ deg$^2$. 

\begin{figure}
\begin{center}
\includegraphics[angle=0, width=240pt]{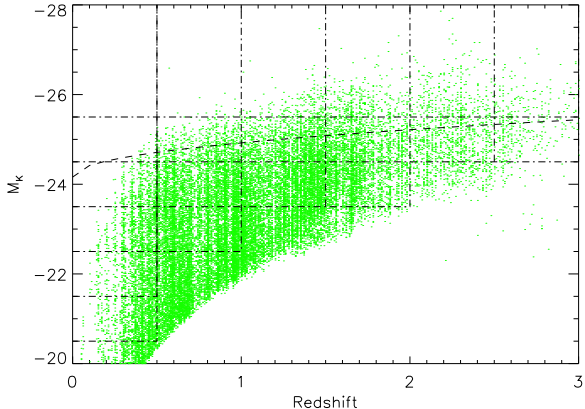}
\caption{The distribution of UDS galaxies used in this work in the photometric
  redshift -- M$_K$ plane. The dot-dashed lines show the sub-sample
  regions for which we compute clustering lengths, provided they
  contain $> 150$ galaxies. Also plotted, as the dashed-line curve, is
  the fit for the evolution of the characteristic luminosity,
  M$^*_K$, of the K-band luminosity function, as given in \protect\cite{Cirasuolo10}.}
\label{Vollim}
\end{center}
\end{figure}

The world-public DR3 catalogue extracted from the K-band image is used
as the basis for our selection, upon which we impose a cut at
$K=21.1$ ($K_{AB}=23$) within $2\arcsec$ diameter apertures. This cut was made to minimise
photometric errors and the number of spurious sources and to
ensure a high level of completeness and robust photometric redshifts
(see below). This limit is fainter than $M^*_K$, the characteristic
luminosity of the K-band luminosity function (\citealt{Cirasuolo10} and
fig. \ref{Vollim}). We are therefore probing the properties of
`normal' galaxies out to $z=3$. Corresponding $2\arcsec$ magnitudes for each
object were extracted directly from the optical, J and H-band images at the position of
the source after a detailed astrometric re-alignment to the UDS K-band
image.

From our combined catalogue we remove stars in the following way:
bright ($K<18.1$) stars were removed by excluding those with K-band aperture
magnitudes within $3\arcsec$ and $2\arcsec$ satisfying
$K_{3\arcsec}-K_{2\arcsec}>-0.14$, where the stars are clearly
separated from the galaxies. The remaining, fainter stars form a
stellar locus in the $(B-z^{\prime})-(z^{\prime}-K)$ plane (as noted
by \citealt{Daddi04}) and can easily be removed by the criterion
$(z^{\prime}-K)<0.3(B-z^{\prime})-0.5$ \citep{Cirasuolo10}. Saturated stars and the surrounding contaminated regions
were carefully masked out during the analysis. These cuts left $33923$
galaxies from which we made our selections, as detailed in the
following section. 

The magnitudes obtained from the images were then used to determine
photometric redshifts (photo-zs) and stellar ages by a $\chi^2$
minimisation over a large suite of model templates, constructed using
\cite{Bruzual03} with a Salpeter initial mass function, and including a
treatment of dust content (see \citealt{Cirasuolo07,Cirasuolo10} for a full account). These templates cover nine
values of the exponential star-formation decay rate ($\tau = 0.1, 0.3,
0.5, 1, 2, 3, 5~$ and $10$ Gyrs, and an instantaneous burst), and
allow stellar ages up to the age of the Universe at any given
redshift. The dust reddening is allowed to vary between $0<A_V<2$,
following the Calzetti law \citep{Calzetti00}, and line of sight
neutral Hydrogen absorption, calculated from \cite{Madau95}, is also
taken into account.

We remove from our catalogue
any source with an unacceptable fit ($\chi^2 > 15$, $4.0\%$ of the
galaxy sample) as they will likely be mis-assigned when we break up
our sample by redshift, and the derived stellar ages are unlikely to
be reliable for such objects. Upon investigation, the majority of these
discarded sources were found to be cross-talk artifacts ($26\%$), QSOs
($36\%$) or the minor members of pairs or mergers ($23\%$). The
majority of the remaining discarded objects were of very low surface brightness
and many of them would not have made it into the
volume-limited sub-samples. The fraction of otherwise useful objects
rejected by this $\chi^2 > 15$ criterion is therefore $<0.6\%$. Rest-frame magnitudes and colours were
computed by integrating the appropriate filter over the
best-fitting template of each object. In this work we make use of the
absolute rest-frame
K-band magnitude and rest-frame $(U-B)$ colour. 

\begin{figure}
\begin{center}
\includegraphics[angle=0, width=240pt]{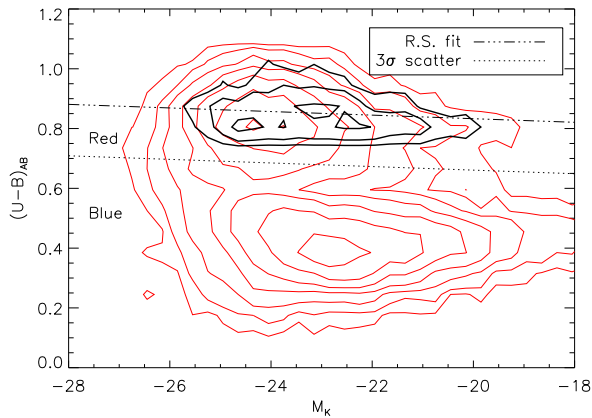}
\caption{Colour-luminosity density plot of the full galaxy sample
  (light, red contours) and of those used to define the red sequence
  (heavy contours). The red sequence was defined by fitting a line of
  the form $U-B = a \times M_{K} + b$ to the galaxies which were best
  fit by a burst template with age $>1~Gyr$. This fit is shown by the
  dot-dashed line, while the dotted line is the $3\sigma$ lower
  envelope. Galaxies redder than the dotted line are defined as `red'
  for the purposes of our sample selections.}
\label{UBsel}
\end{center}
\end{figure}

\subsection{Sample selection}
\label{sample}

In this section we outline our definition of the passive galaxy
sample. At low redshift, passive galaxies form a
well-defined red sequence in a diagram of rest-frame $U-B$ colour
versus absolute magnitude. The red sequence is not exclusively
composed of passive systems, however, and typically contains a $30\%$
contamination from dusty
star-forming galaxies \citep[e.g.][]{Wolf05}. At high redshifts
($z>1$) contamination by dusty star-forming objects becomes increasingly
problematic \citep{Daddi04}. In addition, a simple $U-B$ cut does not
make full use of all the available photometry. For
these reasons, we define our passive sample with two criteria: (a) the
galaxy must be `red' in rest-frame $U-B$ colour and (b) the
best-fitting galaxy template must imply a very low level of residual
star formation. These criteria are further detailed below.

Our rest-frame $U-B$ colours are computed from the best-fitting template used to
find the photometric redshift. The
templates cover two possible star-formation histories: either an instantaneous burst
parameterised by an age, or an exponentially decaying star-formation
rate parameterised by an age and the e-folding time, $\tau$. Hence,
\begin{equation}
SFR_{obs} = SFR_{\rm{initial}} \times e^{-age/\tau}
\label{SFReqn}
\end{equation}
where $SFR_{obs}$ is the SFR at the time of observation.

We wish to define our red sequence initially using the most
conservative and clean sample of passive galaxies drawn from our
catalogue. We construct this sample to be those galaxies which are
simultaneously old (age $> 1$Gyr) and are fit with a burst template,
as these values imply a strong $4000{\rm \AA}~$ break. We then perform a
$\chi^2$ minimisation to fit an equation of the form $U-B_{AB} = a \times
M_K + b$ to these galaxies. The `red' sample is then taken to include
all galaxies within $3\sigma$ scatter of this fit (or redder). In
Figure \ref{UBsel} we plot contours showing where galaxies lie in the
$(U-B)_{AB}$ vs. absolute K-band magnitude plane. Overlaid in black are the
contours for the burst galaxies described above, the associated
best fit (dashed-dotted line) and red/blue selection criterion
(dotted line).

The red sequence
is known to evolve with time, with redder colours at lower
redshifts \citep[e.g.][]{Brammer09}. Accurately accounting for this
evolution with photometric redshifts is problematic. The typical errors
on our U-B colours are $0.15$ magnitudes, which is of the same order
as the expected colour evolution. For simplicity, we choose not to
treat the red sequence in such detail and instead use a fixed colour
selection boundary. We note that
our `red' sample includes part of the green valley, and so we are
unlikely to miss many intrinsically red galaxies at high redshift.

In order to remove the galaxies that are red due to dust rather than
age we make use of the star-formation histories obtained from the
templates. To be selected as a passive galaxy we require the galaxy to
be red, as defined above, older than $1$Gyr and to have a best-fit SFR
(from the photo-z template fits) below some cut-off value (as given by
equation \ref{SFReqn}, where $\tau=0$ for a burst model). By varying
this SFR cut-off criterion we can study how the inclusion of galaxies
with greater recent star formation rates influences the clustering
strengths of the sub-samples. 

We require, for our primary passive sample, a highly conservative
limit to residual star-formation
rate: $SFR_o \le 0.1\%$ of $SFR_{\rm{initial}}$ (in addition to being
red in rest-frame $U-B$ and with a stellar age $>1$Gyr). We choose
this strict limit to minimise the contamination by dusty star-forming
objects whilst retaining sufficient objects with which to compute
meaningful clustering lengths. There are
well-known degeneracies between age, metallicity and dust content. We
cannot completely overcome them, even with the multiple bands of deep
photometry at our disposal. However, these parameters imply a strong break in
the galaxy SED and hence only the most extreme dusty contaminants will
be included. The number of objects satisfying these selection criteria
is 3991 ($33\%$ of the `red' sample). The vast majority of passive
galaxies selected in this way also lie within the `pERO' colour
selection boundaries (see section \ref{disc}).

We choose, as a baseline against which to compare our passive sample,
an actively star-forming sample with a single criterion: $SFR_o > 10\%$ of
$SFR_{\rm{initial}}$. This star-forming sample consists of 22856
objects of which $91\%$ are `blue' according to our (U-B) criterion and
$9\%$ are `red'. We then construct three further passive samples by
relaxing the SFR criterion used in our primary passive sample to include galaxies with
$SFR_o \le 0.5\%$, $SFR_o \le 1\%$ and $SFR_o \le 10\%$ of
$SFR_{\rm{initial}}$. 
In each of these three additional samples the
requirements for red $U-B$ colour and stellar age are kept the same as
the primary sample. The number of
objects in each of these expanded passive samples are 6111, 7677 and
10032 respectively. We will occasionally refer to these samples in the
text by the SFR limit imposed during selection.

\section[]{Clustering properties}

The real-space correlation length of a sample of galaxies is
intimately related to the mass of dark matter halos in which they are
found. The value of this clustering scale-length is therefore of great
interest in the study of galaxy formation and evolution. In this work we determine the angular
correlation function of the galaxies as they are seen in projection,
and then use the redshift distribution to deproject to find the
real-space scale length ($r_0$). This process is detailed below.

The 2-point angular correlation function, $w(\theta)$ is defined by the joint
probability of finding two galaxies in solid angle elements $\delta\Omega_1$
and $\delta\Omega_2$ at a given separation \citep{Peebles80},
\begin{equation}
\delta P = n^2 \delta\Omega_1 \delta\Omega_2 (1 + w(\theta_{12})).
\end{equation}

To estimate the correlation function we use the estimator of \cite{Landy93},

\begin{equation}
w(\theta) = \frac{DD - 2DR + RR}{RR}
\end{equation}

\noindent where DD, DR and RR are the counts of data-data, data-random and
random-random pairs respectively at angular separation $\theta$,
normalised by the total number possible. Although this estimator is
relatively robust against systematic errors, there remains a small bias due
to the finite field size. This bias is corrected by an integral
constraint, constant C. We follow the method of \cite{Roche99} by using
the random--random counts to estimate the size of this bias:

\begin{equation}
C=\frac{\Sigma N_{RR}(\theta)\theta^{-0.8}}{\Sigma N_{RR}(\theta)},
\end{equation}

\noindent where the sums extend to the largest separations within the field.

The angular correlation functions are fit by $\chi^2$ minimisation
with a power law of fixed slope, $\delta= -0.8$. The fit is made over
mid to large scales to avoid biases arising from multiple halo
occupation. The small-scale limit corresponds to
$0.3h^{-1}$Mpc\footnote{This value was chosen to maximise the fitting
  range, though we have also tried $0.5h^{-1}$Mpc and $1.0h^{-1}$Mpc,
  finding consistent results.}, where
previous $w(\theta)$ measurements of galaxies over $1<z<2$ begin to diverge
from the large scale power law (H08). The large scale cut off is at
$0.2$ degrees where survey area limits
our measurements. This method is the same as that presented in
H08. Correlation function uncertainties are determined using bootstrap
resampling, which are typically 2-3 times the Poisson errors.

The real space clustering and projected clustering are linked by the
relativistic Limber equation \citep{Limber54}.  If the redshift
distribution of a sample is known, the Limber equation can be inverted and the
correlation length, $r_0$, can be calculated in a robust manner \citep{Peebles80,
Magliocchetti99,Roche03}. Following \cite{Magliocchetti99} and
adjusting to take into account our known photometric redshift
distributions we determine r$_0$ as

\begin{equation}
r_0 = \left[\frac{cA}{H_0H_{\gamma}}\left(\frac{(\int^{\infty}_0~n(z)dz)^2}{\int^{\infty}_0~n(z)^2x(z)^{1-\gamma}P(\Omega_0,z)F(z)dz}\right)\right]^{1/\gamma}
\end{equation}

\noindent where $n(z)$ is the redshift distribution of the sub-sample
and all other symbols have the same meanings as in
\cite{Magliocchetti99}.

Below we present the resulting measurements for $r_0$ as a function of
redshift and K-band luminosity. This is first determined for all
galaxies and then further subdivided by colour and star-formation history.

\subsection{The global evolution of clustering}

\begin{figure}
\begin{center}
\includegraphics[angle=0, width=240pt]{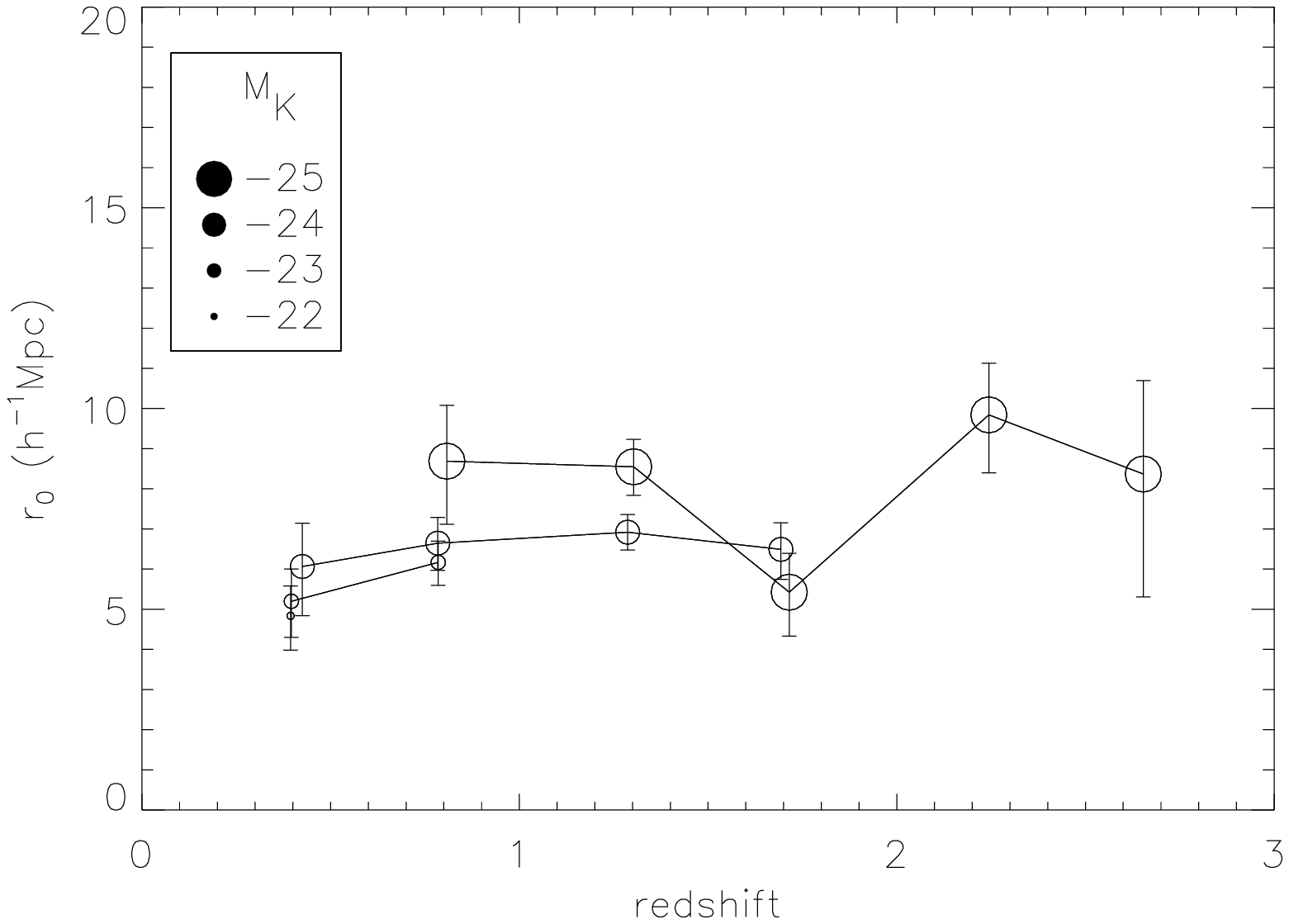}
\includegraphics[angle=0, width=240pt]{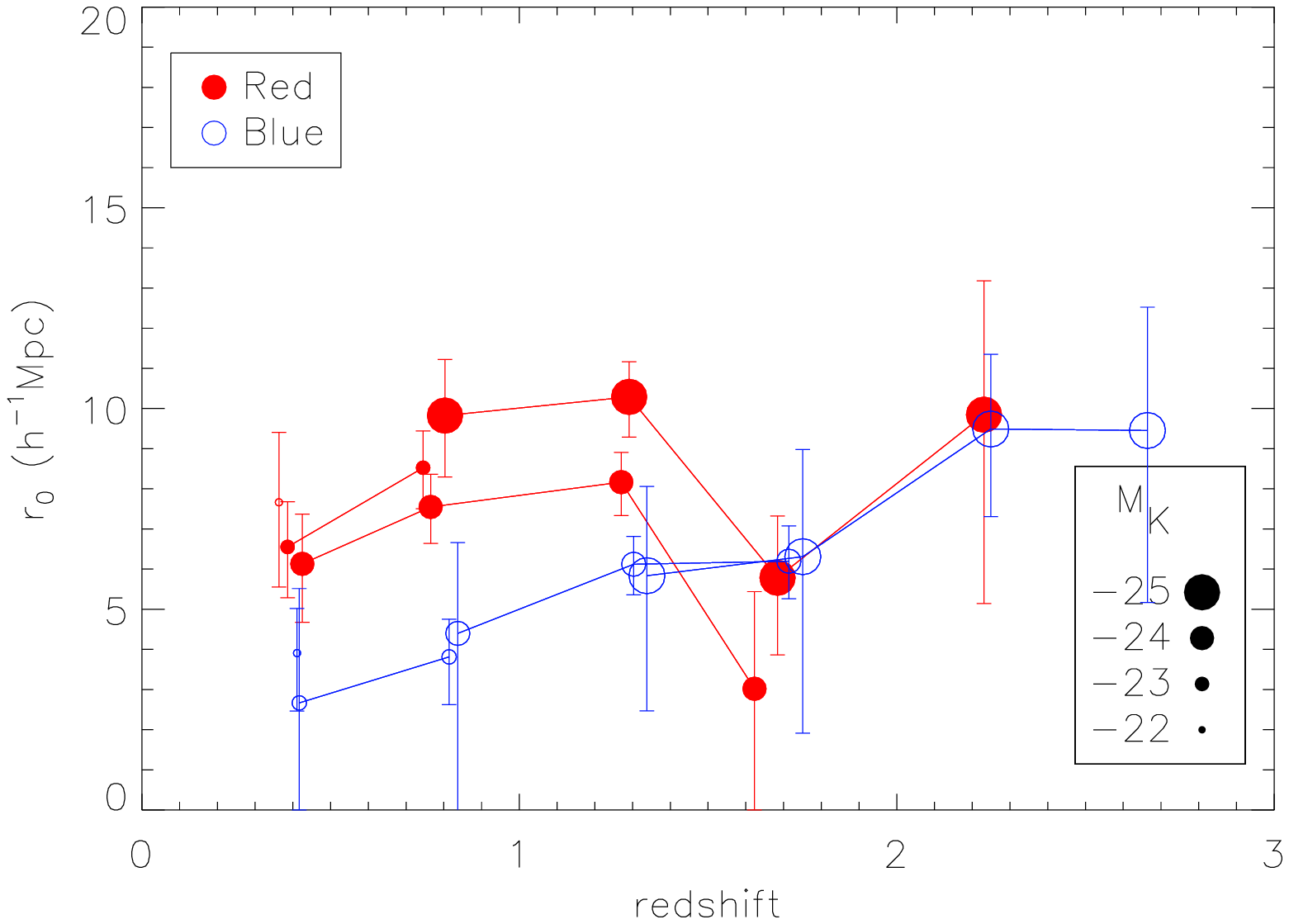}
\caption{{\bf{\em Upper.}} The global evolution of clustering in the UDS
  field. Galaxies are split into the volume-limited sub-samples shown in Figure
  \ref{Vollim} and the real-space correlation lengths are computed
  (see text). The evolution of clustering strength with redshift is
  mild for most luminosity ranges. {\bf{\em Lower.}} The sub-samples are further separated into
red and blue in rest-frame $(U-B)$ colour and the clustering computed
once again. Globally, red galaxies are slightly more strongly
clustered than blue galaxies. The increase in $r_0$ found for bright
galaxies at low-redshift is clearly seen to be dominated by red
galaxies.}
\label{fullUB}
\end{center}
\end{figure}

Using the method outlined above we computed the galaxy clustering scale
lengths, $r_0$, for each of the volume limited sub-samples
shown in Figure \ref{Vollim}, provided that each subsample contained at
least 150 galaxies. These measurements are shown in Figure
\ref{fullUB}. The first thing of note in these measurements is the lack of
strong evolution of $r_0$ with redshift at a given $M_K$. Furthermore, there is very
little luminosity segregation, consistent with the findings of
\cite{Coil08}, although
the most luminous galaxies are slightly more strongly clustered at
$z\sim1$.

\subsection{Colour and star-formation rate dependant clustering}

\begin{figure*}
\begin{minipage}{150mm}
\begin{center}
\includegraphics[angle=0, width=\textwidth]{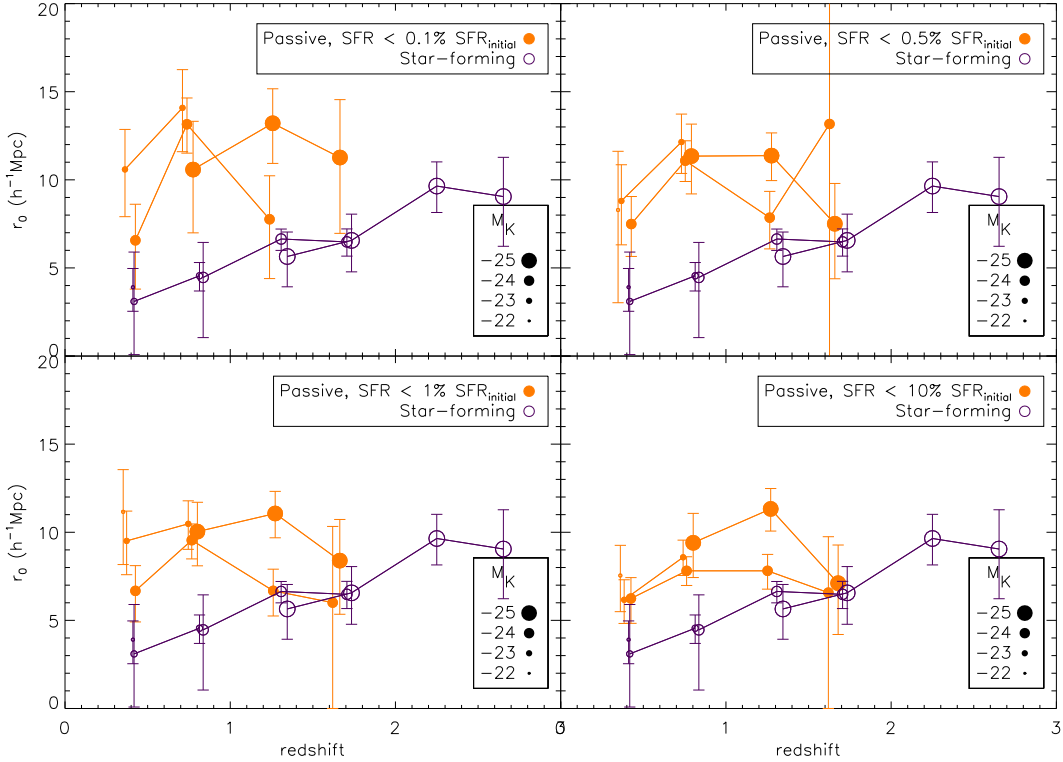}
\caption{Evolution of the clustering strengths with redshift for our four 
  passive samples with the single comparison star-forming sample. 
With the most conservative passive definition
  (requiring red rest-frame $U-B$ colour, stellar age $>1$ Gyr
  and $SFR_o\le0.1\%~SFR_{{\rm initial}}$), the passive galaxy
  sub-samples are significantly more strongly clustered than the star-forming
  sub-samples for $z<1.5$. Above $z=2$ the number of
  passive galaxies becomes too small to make robust clustering measurements. 
As the passive galaxy selection criterion of SFR  
(given in each panel) is relaxed, the
  clustering strengths of star-forming and passive sub-samples become
  more similar. Each point is plotted at the mean redshift of the
  sub-sample it represents.}
\label{result}
\end{center}
\end{minipage}
\end{figure*}

\begin{figure}
\begin{center}
\includegraphics[angle=0, width=240pt]{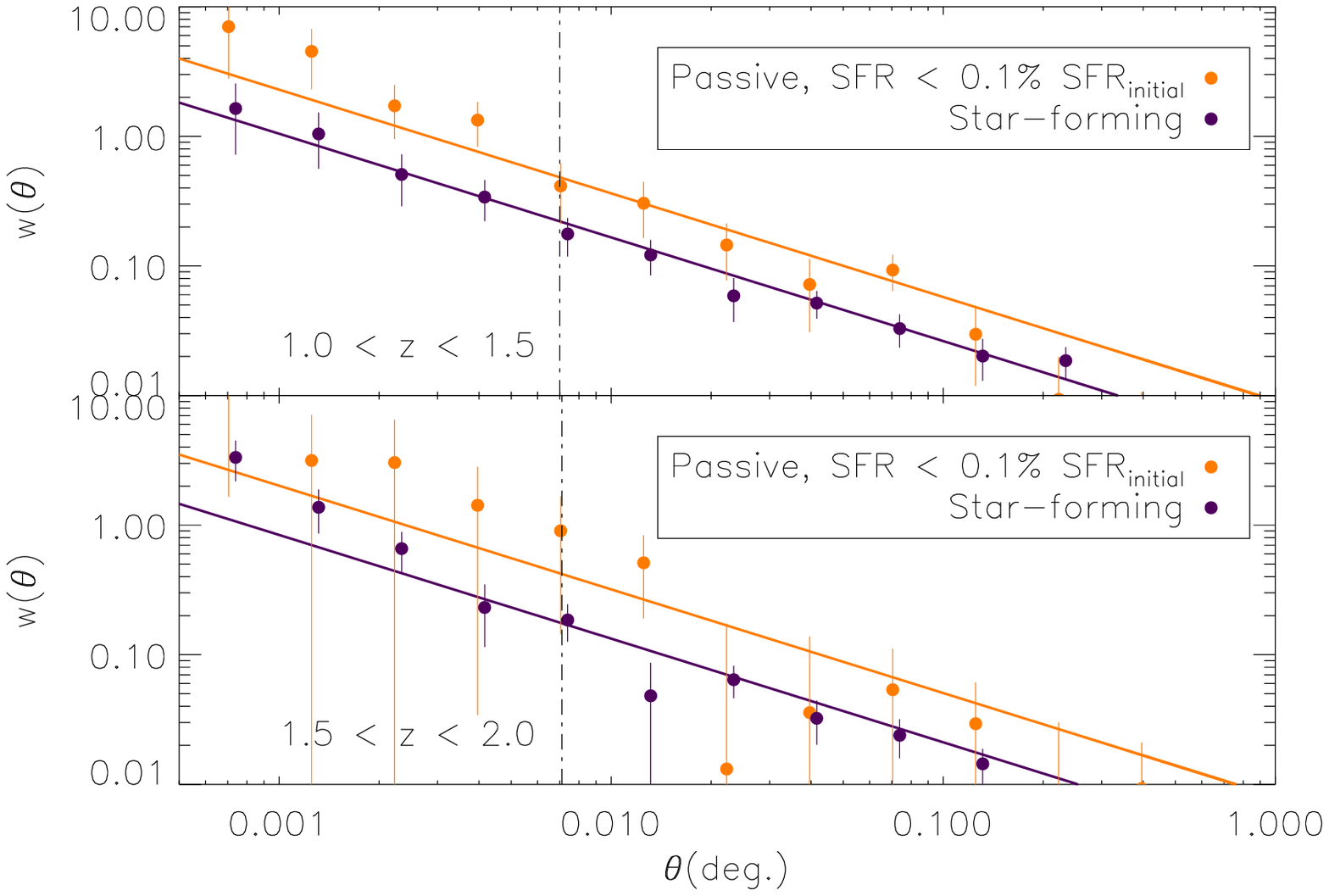}
\caption{Angular clustering measurements for conservative passive
  galaxies ($SFR_o \le 0.1\%~SFR_{\rm{initial}}$) and star-forming
  galaxies within the redshift ranges $1<z\le1.5$ (upper) and
  $1.5<z\le2$ (lower) with the star-forming points off set for
  clarity. Subsamples in luminosity have been combined for
  illustrative purposes. Solid lines show the best power-law fits, using a constant
  slope of $-0.8$, over large-scales: $0.3~h^{-1}{\rm Mpc~ (Dashed line)} - 0.2~$deg. In
  both cases the passive samples are more strongly clustered,
  consistent with the measurements of pBzK and sBzK-selected
  galaxies (e.g. H08). Errors are determined from a bootstrap analysis.}
\label{wthplot}
\end{center}
\end{figure}

In much of the literature concerning the evolution of galaxy
clustering, populations are separated by colour, rather than
star-formation properties. Rest-frame colours are simpler
to compute and therefore allow more straightforward comparisons with
previous work. In Figure \ref{fullUB} we again compute $r_0$ values
for the subsamples in Figure \ref{Vollim}, but this
time we also separate red and blue galaxies, using our rest-frame
$(U-B)$ colours defined in Section \ref{sample} (Figure \ref{UBsel}).

At $z<1$, red galaxies are significantly more clustered than blue
galaxies, similar to the result of \cite{Coil08} at $z\simeq1$. This 
difference is particularly pronounced for the brightest
subsamples, where the strong clustering observed in the full sample is
clearly a result of the clustering of red galaxies.
Above $z=1.5$ the clustering strengths of the blue and red subsamples
appear to converge, indicating that this may be the principle epoch at
which the red sequence is populated. The red population is
inhomogeneous, however, containing both passive and dusty star-forming
objects. We therefore use our template fits to separate the passive
and star-forming galaxies (Section \ref{sample}). 

We plot the deprojected clustering lengths of our passive and
star-forming samples in Figure \ref{result} with the SFR limits
described in section \ref{sample} shown in each panel. For the most
conservative passive sample (SFR $\le 0.1\%$) the clustering strengths of the
passively-selected galaxies are greater than those of the equivalent
star-forming galaxies for almost all sub-samples up to $z=1.5$. Our
result, shown in the upper left panel of Figure \ref{result}, 
shows that the clustering length is more strongly dependent on passivity than on
luminosity, with very little luminosity segregation in the
star-forming sample. This difference continues to at least the
median redshift of BzK-selected galaxies \citep{Hartley08}. Above
$z\sim1.5$, however, the respective clustering
strengths of passive and star-forming galaxies appear to converge. 
Further survey data will be required to confirm
this finding, which is discussed in section \ref{disc}. As we
relax the SFR requirement to allow galaxies with higher residual
star-formation rates into the passive selection, we find that the
strength of clustering of the passive sub-samples decreases, which
confirms that truly passive galaxies are more strongly clustered among
the `red' population.

For illustration, in Figure \ref{wthplot} we show the angular clustering measurements
for passive and star-forming galaxies over all sub-samples 
in the two redshift ranges, $1<z\le1.5$ and $1.5<z\le2$. The
passive samples use our conservative definition
($SFR_o\le0.1\%~SFR_{{\rm initial}}$) described in section
\ref{sample}. The solid lines are the best power-law fits and, in
agreement with our earlier measurements for BzK-selected galaxies (H08), the
passive galaxies are found to cluster more strongly.

\subsection{Comparison with modelled halo masses}

We use the formalism given in \cite{Mo02} to find the clustering
strengths of dark matter halos of given masses as a function of redshift. These
models gather together previous work
\citep{Press74,Bond91,Mo96,Jing98,Sheth01} and use the ellipsoidal
collapse model which has been calibrated against dark-matter only
N-body simulations. Their aim is to provide the abundances and
clustering of dark matter halos as functions of redshift and mass
across a wide dynamic range.

\begin{figure}
\begin{center}
\includegraphics[angle=0, width=240pt]{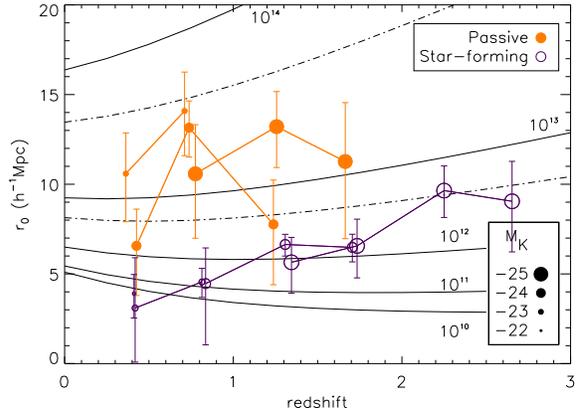}
\caption{The clustering of our conservative passive sample and
  star-forming sample (symbols have the same meanings as in Figure
  \protect\ref{result}) are compared with the clustering predictions
  of dark matter halos from \protect\cite{Mo02}. Lines of constant
  halo mass (in $M_{\odot}$) are shown, with additional dashed-dotted
  lines for $5\times10^{13}~M_{\odot}$ (upper) and
  $5\times10^{12}~M_{\odot}$ (lower). Downsizing in the star-forming
  population is evident with the same luminosity galaxies found in
  less massive halos towards z=0. The same cannot be said of the
  passive galaxies, which typically occupy halos of mass $M\ge10^{13}M_{\odot}$.}
\label{MWplot}
\end{center}
\end{figure}

In Figure \ref{MWplot} we plot these models for halos of mass
$10^{10}M_{\odot}$ to $10^{14}M_{\odot}$ together with the results
from our conservative passive sample (see section \ref{sample}) and star-forming sample. At $z<2$, star-forming
galaxies generally occupy halos of mass $10^{12}M_{\odot}$ or smaller,
while passive galaxies occupy halos of mass $10^{13}M_{\odot}$ to
$5\times10^{13}M_{\odot}$. The general
trend for star-forming galaxies is for galaxies of a given luminosity
to occupy halos of greater mass at higher redshift. This is a
manifestation of halo downsizing (\citealt{Cowie96}; Foucaud et al. 2010,
in press). This downsizing in host halo mass for star-forming
galaxies appears to extend beyond z=2. At these very early times
bright star-forming galaxies are hosted by massive dark matter
halos ($M\sim5\times10^{12}M_{\odot}$). Presumably these are among the
progenitors of low redshift group-dominant galaxies.

The passive galaxies occupy halos up to an order of magnitude more
massive than the star-forming galaxies. In contrast to the star-forming
galaxies, the brightest passive galaxies appear to show constant
clustering strength over the
range $0.5<z<1.5$. If confirmed by future survey data, 
this result suggests that bright passive 
galaxies are hosted by similar mass halos across this redshift range.

\section[]{Discussion}
\label{disc}

We have shown that passive galaxies, defined by fitting templates over
11 bands from the UV to {\it Spitzer}'s IRAC bands, are more strongly
clustered than star-forming galaxies at $z< 1.5$. 
Clustering strength is intimately linked with the minimum mass of dark
matter halo that can host a galaxy of a given type. Passive
galaxies therefore on average occupy more massive dark matter halos
than their star-forming counterparts. These results indicate that at
$z < 1.5$ the passivity of a galaxy sample is a strong indicator of the
typical mass of dark matter halo that host them.

The K-band luminosity is a reasonable proxy for the stellar mass of a
galaxy, though galaxies may dim by approximately one magnitude after
they stop forming stars \citep[e.g][]{Lilly84,Cowie96}. Even taking this
dimming into account, the conservative passive sub-samples
 ($SFR \le 0.1\%~ {\rm SFR}_{{\rm initial}}$) are typically more strongly clustered
than the star-forming samples for a given stellar mass.

Furthermore, there is an absence of a strong luminosity dependence on
the clustering strength of the star-forming samples. 
We have therefore found that of the two effects that we set out to
disentangle, passivity and luminosity, passivity is the more
significant indicator of clustering strength at $z<1.5$. However, we
note that we are unable to probe the most extreme high luminosity galaxies
(M$_K<-26$) due to our limited survey area.

Above $z=1.5$ we are unable to significantly distinguish star-forming
and passive galaxy clustering. The correlation lengths of passive and
star-forming galaxies apparently converge at this epoch. If this
behaviour is confirmed it would suggest that the epoch $z\sim2$ is the
epoch in which the passive and star-forming samples are first becoming
distinct. Hence, it is likely to be the major epoch at which the red
sequence is being populated. This finding makes the further
study of the $1.5<z<2.5$ range critical and we intend to return to and
improve upon this work as the UDS data push deeper and spectroscopic
samples become available.

In the hierarchical formation of structure the first halos to collapse
are those which eventually merge to form the most massive halos
at low redshifts. Our result then points towards a time sequence:
passive galaxies formed earlier in those first halos while those of
similar stellar mass, but still forming stars, developed in lower mass
halos that collapsed later. Galaxy and halo evolution is accelerated
at the earliest epochs with respect to the present day. The galaxies
formed at these strongly clustered, high density peaks of the matter
distribution are therefore likely to become fully evolved, passive
galaxies more rapidly than the general population. Discounting mergers, the stellar mass of a
galaxy is limited by the available gas reservoir so a passive galaxy
could have formed significantly earlier than a star-forming galaxy,
but end up with very similar stellar masses at the epoch of
observation. Their respective halos, however, will have built-up mass
since they first collapsed and so those of the earlier formers will be
more massive. In this way the difference in clustering strength is a
natural result of hierarchical mass assembly in halos and downsizing
in galaxies (c.f. Foucaud et al. 2010, in press).

In addition to the relative differences in clustering strength at a
fixed redshift, we also find a potential difference in how the passive
and star-forming samples evolve with redshift. The star-forming
samples follow the behaviour we would expect as a result of the
downsizing scenario for galaxy evolution. Sub-samples of a given
K-band luminosity (stellar mass) taken from this population exhibit a
decline in clustering strength towards $z=0$. The most luminous 
{\em  passive} galaxies show tentative evidence for constant
clustering strength across $0.5<z<1.5$. A constant value of r$_0$ of 
such magnitude indicates that the
hosting dark matter halos are no less massive at lower
redshift.

Direct comparison with the results of previous work is extremely
difficult as the UDS is currently unique and this study is the first
of its type utilising K-band selection. Qualitatively, our results
agree very well with the findings of \cite{Coil08}, \cite{McCracken08} and
\cite{Meneux06}: each of these studies finds a longer clustering
lengths for red or passive galaxies over blue or star-forming galaxies
below $z=1$. We furthermore agree with previous works that any
luminosity dependence on clustering in the sample is minimal in
comparison with that of passivity over the range of galaxy properties
that we have investigated.

\cite{McCracken08} find that at intermediate redshifts ($0.5<z<1$),
galaxies fit by an early-type template possibly have a inverse
luminosity dependence, with less luminous galaxies having slightly
longer correlation lengths. This behaviour is well known at low
redshift, where passive dwarf galaxies exhibit strong clustering and
are found to be associated with clusters of galaxies
\citep{Conselice03,Zehavi05}. Our measurements for red galaxies at
$z<1$ are not inconsistent with these findings.

\subsection{Discussion of uncertainties}

Throughout this work there are uncertainties that are extremely
difficult to quantify. Though dust attenuation is taken into account
during template fitting, there is a well-established degeneracy
between the stellar age and dust content. Even with the wide range of
photometry available, we cannot fully overcome this degeneracy. 
In addition, the typical redshift uncertainties
are $\sim0.05(1+z)$ and typical uncertainties in U-B colour are
$0.15$ magnitudes. We use the ERO definitions of \cite{Pozzetti00} to examine
whether our passive selection method is robust. Each of the passive
objects with best-fit redshifts within the range $1<z\le2$ meets the
$i-K>4$ (Vega) ERO criterion,
with 1423 of the 1457 objects lying in the passive ERO (pERO)
region, defined by i,J,K colours. In addition, when plotted in the $(B-z)_{AB} - (z-K)_{AB}$ plane, our
conservative passive sample populates the areas in which we would
expect to find passive galaxies (see the Appendix, Figure \ref{BzKsel} and
\citealt{Lane07}). We are therefore confident that the SED-fit star formation
properties we use are relatively robust, in comparison with previous
broad band techniques. 

Even with robust colours and selection criteria, the photometric redshift technique can introduce
uncertainties into our measurements by causing a galaxy to be assigned
to the wrong sub-sample in redshift. Passive galaxies have a strong break implied by our
selection process and hence the photometric redshift determination is
more reliable for these galaxies than for star-forming
galaxies. There is an ongoing ESO Large Programme, the UDSz survey, to obtain
$\sim4000$ redshifts of galaxies at $z>1$ in the UDS field. Early
indications suggest that the photometric redshifts for our passive
galaxies are indeed highly reliable, with error estimates of
$\delta_z\sim0.02(1+z)$ (though at the time of writing the majority of
these are at $z<1.75$). 

Photometric redshifts
are particularly difficult to obtain at $z>1.5$. It is precisely
beyond this point that the red and blue samples converge, due in part
to a decrease in the red galaxy clustering. On the other hand, the
blue and star-forming samples' clustering strengths increase over this
range. Sub-sample misassignment, particularly in redshift, will
tend to dilute the clustering signal. For the observed convergence to 
be driven by errors in the photometric
redshifts would therefore require the clustering signal from our red
sample to be washed out, while that from the blue sample is not. This
in turn would require the redshifts of our blue and
star-forming galaxies to be much more accurate than our red galaxies. We
suggest that such a case is highly unlikely, but await a substantial
sample of redshifts to confirm this.

Of greater concern is the
possibility that due to photometric redshift errors, star-forming
galaxies are more likely to be assigned
to the wrong redshift bins. This would dilute the clustering of the
blue galaxies and artificially enhance the difference between passive
and star-forming samples. If dilution due to redshift errors is significant
we might expect that our clustering measurements for star-forming
galaxies are smaller than other studies of star-forming galaxies at
these redshifts. In H08 the clustering strength of sBzK-selected galaxies for a sample
of median redshift $\sim1.5$ 
cut at $K=23$ was found to be $\sim6.75~h^{-1}$Mpc. This value is
similar to the star-forming sub-samples of similar redshifts presented
here. Although selections in other bands are not directly comparable
we have found that clustering strength is only weakly dependent on
K-band luminosity. Any moderately bright star-forming sample can
therefore be used as a comparison. \cite{Coil08} found that star-forming
galaxies at $z=1$ have modest clustering strengths of $\sim
4h^{-1}~$Mpc, with similar values found by \cite{Adelberger04}
(BM/BX-selected galaxies, $z=1$) and \cite{McCracken08} ($z=0.6$). The $z=1$,
UV-selected galaxies in \cite{Heinis07} have a slightly larger
correlation length ($4.1 h^{-1}{\rm Mpc}<r_0<5.5 h^{-1}{\rm Mpc}$),
similar to those we report. At similar redshifts we find slightly
longer ($3 h^{-1}{\rm Mpc}<r_0<5 h^{-1}{\rm Mpc}$), but consistent, 
correlation lengths compared with most other studies.

A second method we can use to test this potential issue was also used
in \cite{Magliocchetti99}. \cite{Magliocchetti99} noted that there is
a systematic error introduced into the r$_0$ values when it is assumed
that the redshift distribution is accurately known. They take account
of this error by assuming the errors in the redshifts of their
sub-samples are Gaussian random and apply a co-efficient,
$[(12\sigma^2/\Delta z^2)+1]^{1/2\gamma}$, to broaden the redshift
distribution. In this work modelling the individual $n(z)$ bins as
top-hat functions would be inaccurate and so we have not attempted to adjust
our $r_0$ measurements in this way. However, this method can still
provide an estimate of whether our result is driven by redshift errors.
Each galaxy has an associated error in redshift, derived from the
$\chi^2$ distribution, which we assume here can be represented by a simple Gaussian. In this way
each galaxy in a sub-sample has a probability of being at a given
redshift and these individual probabilities are then binned and summed
for all galaxies within a sub-sample. The resulting distribution is
then used in place of the original n(z) during deprojection. We find
that the $r_0$ values increase, as expected from a broader redshift
distribution. However, the difference in clustering between the
star-forming and passive sub-samples remains, though the difference in
implied halo masses is slightly reduced. 

This method is very much a
simplification and the errors associated with the photometric
redshifts are likely to be much more complex. In particular the errors
on the star-forming galaxies' redshifts may deviate significantly from
a Gaussian. Photometric redshift codes are known to favour certain
redshifts, introducing artificial clumping in redshift space. The spectroscopic 
redshifts obtained from the UDSz survey will be
vital in understanding possible biases arising from using photometric redshifts.
We conclude from our current tests, however, that errors in
the redshift determination are unlikely to be responsible for the results that we
have found.

A further, and potentially important source of error, is that of
cosmic sampling variance. In contrast to H08, we have chosen not to
take the effects of sampling variance into account. The method
employed in H08 consisted of splitting the field into 4 and computing
a measurement for each sub-field. This method provides a estimate of
the sample variance for a field only one quarter of the size of the
original. The ideal method for estimating the sample variance is to
repeat these measurements for comparable, independent fields. To
facilitate such work in the future we re-define our samples, using
simple photometric selection in the Appendix. However, we also note
that the clustering of sub-samples at different redshifts are largely
uncorrelated, yet we find consistent differences between passive and
star-forming galaxies for subsamples with
$z<1.5$. We suggest that it is unlikely that we would find such
consistent disparity in the clustering results due to cosmic sampling
variance alone.

\section[]{Conclusions}

We have used the UDS, the deepest contiguous near infrared survey
currently available over an area of $\sim 0.7~ {\rm deg}^2$ to
investigate the clustering strengths of galaxies over the range
$0<z<3$. Best fitting templates,
including a dust contribution, have been measured for each
galaxy. From these templates the rest frame U-B colour, absolute
K-band magnitude, stellar age and e-folding time of star-formation,
$\tau$, are derived. These quantities are then used to define `red'
and `blue' galaxies, a
conservative passive sample and active star-forming sample. 

Each of these samples are
then further sub-divided by redshift and K-band luminosity. Angular
correlation function measurements are made for each of these
sub-samples and power laws are fit over the scales
$\theta_{lim}<\theta<0.2$ degrees, where $\theta_{lim}$ corresponds to
$0.3h^{-1}$Mpc. This limited fitting range is used so as to avoid the
bias caused by multiple halo occupation (on small scales) and to avoid the scales where the
correlation function and integral constraint correction become similar. 
Using photometric redshift distributions, the angular
correlation functions are deprojected to find real space correlation
lengths.

We find that the conservatively-defined passive sample is more strongly
clustered than the star-forming sample for all luminosities at
$z<1.5$. Above this redshift the clustering strengths appear to
converge, though we would require greater depth and statistics to make
a firm conclusion. Furthermore, by relaxing our strict passive galaxy
selection we show that passive galaxies are the most strongly
clustered of the `red' galaxies.

Clustering strength is intimately related to the
typical mass of dark matter halo hosting a sample, and K-band
luminosity is a reasonable proxy for the stellar mass of a galaxy. We
have therefore shown that passive galaxies of a given stellar mass are
hosted by more massive halos at $z<1.5$. At higher redshift the
convergence of clustering strengths suggests that the red sequence is
in the early stages of being populated.

We have also studied how the clustering of passive and star-forming
sub-samples compares with those of dark matter halos of varying
mass. The clustering strengths of star-forming galaxies decline
towards $z=0$ for a given K-band luminosity (stellar mass), and hence are
typically found in less massive dark matter halos. This finding is
consistent with the wealth of evidence supporting downsizing.

The UDS imaging project is ongoing, and expected to gain at least 1
magnitude in J, H, K depth by 2012, and potentially substantially
deeper. This will enable us to extend this work to higher redshift and
lower luminosities. In addition, our ongoing spectroscopic programme
(UDSz) will provide several thousand spectra over the redshift range
probed by this analysis. This will enable a variety of more detailed
studies, in addition to substantially improving the reliability of
photometric redshifts. The addition of similar deep fields from
upcoming surveys (e.g. UltraVISTA) will also allow robust estimates of
the sample variance. We therefore anticipate major progress in our
understanding of the galaxy populations at the crucial epoch when the
red sequence first becomes established. 

\section*{Acknowledgments}
WGH would like to extend his thanks to Alfonso Arag\'on-Salamanca for
an extremely useful discussion during the preparation of this
work. WGH and SF would also like to acknowledge the support of the
STFC during the preparation of this work. IRS acknowledges support
from STFC. RJM acknowledges Royal
Society funding through the award of a University Research
Fellowship. JSD acknowledges the support of the Royal Society through
a Wolfson Research Merit award. Finally, we would like to thank the
anonymous referee for their very thorough reading of our work and
their insightful comments.

\section*{Appendix}

In this appendix we re-visit the 2-colour selection of
\cite{Daddi04}. We aim to derive a series of simple boundaries which
can be used to approximate the passive samples used earlier in this work.
Our purpose here is to facilitate the reproduction of
our results in complementary fields rather than create a robust
  definition for a passive sample. We choose the $(B-z^{\prime})_{AB} -
(z^{\prime}-K)_{AB}$ plane because the pBzK selection of \cite{Daddi04} is well
    understood,
and also because of the track identified in
\cite{Lane07}. This track has colours consistent with the SED of an E/S0
type galaxy,
 evolved through redshift, and is well separated from the rest of the
 distribution. It suggests that it may be possible to separate passive
 and star-forming galaxies over all redshifts with relative ease.

The use of such simple photometric sample
definitions may be preferable to full star-formation history fitting when
comparing different data sets. In minimising the complexity of
selection, the intrinsic differences in the samples should be more
apparent and hence a better estimate for cosmic sample variance can be
obtained. The colour boundaries outlined below are defined using the
Subaru and UKIRT filters. Conversion to other filter sets
can be made with reference to \cite{Hewett06} and \cite{Furusawa08}.

\subsection*{Samples in the $(B-z^{\prime}) - (z^{\prime}-K)$ plane}
\label{border}

\begin{figure}
\begin{center}
\includegraphics[angle=0, width=240pt]{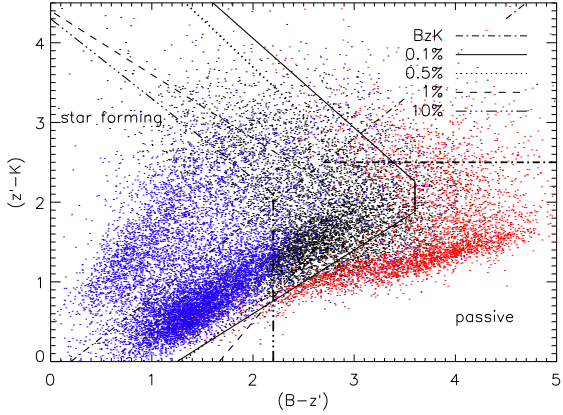}
\caption{$(B-z^{\prime})_{AB}-(z^{\prime}-K)_{AB}$ colour-colour plot for
  actively star-forming sample (blue points), our
  `conservative' passive galaxies (red) and those which fall
  between these criteria (black). 
Also shown are the commonly
  used BzK selection criteria of \protect\cite{Daddi04} and the
  borders defined in the appendix. 
Only half of the star-forming points are plotted for clarity.}
\label{BzKsel}
\end{center}
\end{figure}

The plane is split into cells, $0.2\times0.2$ in colour, and within each
cell the fraction of galaxies that are in the passive sample is
computed. Boundaries are chosen to provide a simple selection technique which includes those cells with $>50\%$
passive galaxies. These borders are shown in Figure \ref{BzKsel}
together with the conservative passive sample and star-forming samples
used in the main body of this paper. The boundaries for each sample are:
\begin{eqnarray}
 &&(z^{\prime}-K)_{AB}<\alpha(B-z^{\prime})_{AB} + \beta \nonumber \\
 &{\rm or}&(B-z^{\prime})_{AB}>\gamma \nonumber \\
 &{\rm or}&(z^{\prime}-K)_{AB}>\delta(B-z^{\prime})_{AB} + \epsilon;
\end{eqnarray}

\noindent where the values for $\alpha, \beta, \gamma, \delta~ {\rm
  and}~ \epsilon$ are given in table \ref{co-effs}.

\begin{table}
\caption{Values for the boundary criteria defined in the text.}
\begin{tabular} {|l|l|l|l|l|}
\hline
Co-eff&$0.1\% SFR_{i}$&$0.5\% SFR_{i}$&$1\% SFR_{i}$&$10\% SFR_{i}$\\
\hline
$\alpha$&0.8&1.0&1.5&...\\
$\beta$&-1.0&-1.3&-2.5&...\\
$\gamma$&3.6&...&...&2.2\\
$\delta$&-1.4&-1.2&-0.8&-1.0\\
$\epsilon$&7.3&6.1&4.4&4.3\\
\hline
\end{tabular}
\label{co-effs}
\end{table}

These 2-colour-defined passive samples include all galaxies in the
relevant region regardless of their best-fit age, U-B colour or
SFR. The number of objects in each sample are 5438, 7287, 8986
and 12242 respectively, and the number of objects in the star-forming
sample is 20645.

\begin{figure*}
\begin{minipage}{150mm}
\begin{center}
\includegraphics[angle=0, width=\textwidth]{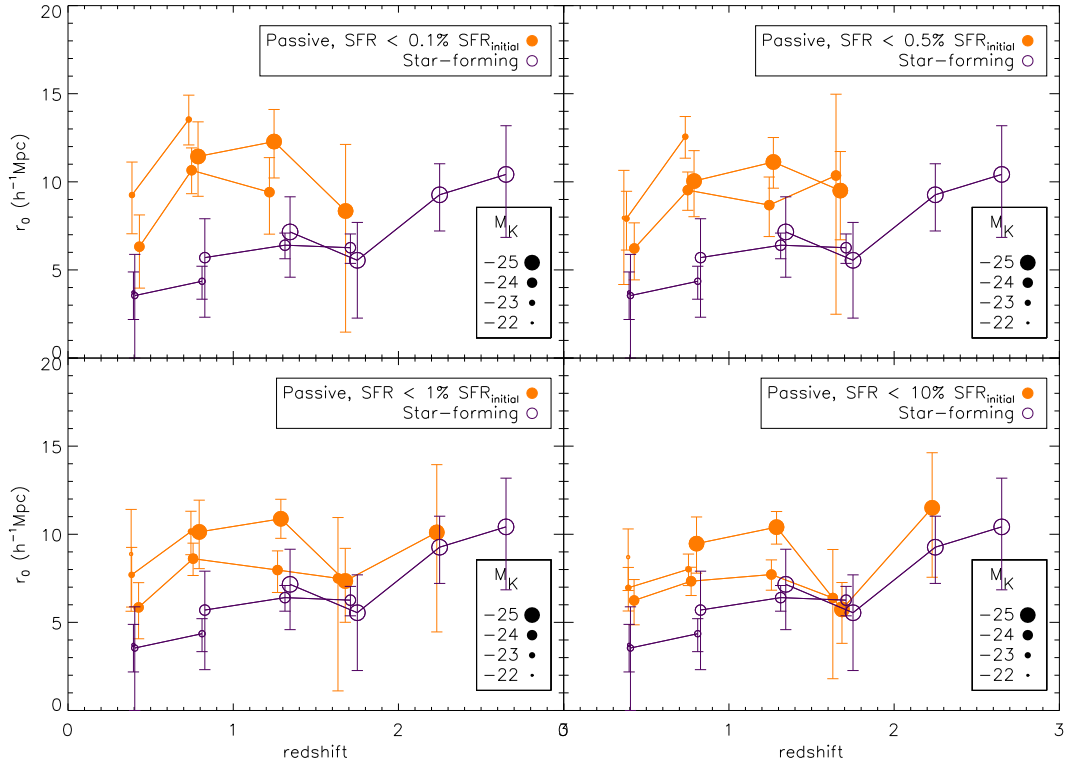}
\caption{Clustering strengths of the 2-colour-defined galaxies. The
  symbols have the same meanings as in Figure \ref{result}. The values
  for r$_0$ are largely consistent with those in Figure
  \ref{result}. Use of the boundaries defined for these samples should
  therefore yield measurements suitable for comparison with our main
  results.}
\label{board}
\end{center}
\end{minipage}
\end{figure*}

The clustering properties of the these galaxy samples were
treated in the same way as those in the main body of this paper. In
each case the properties of galaxies selected by the
photometrically-defined
 borders, shown in Figure \ref{BzKsel}, reflect those of the
intended population: the majority of sub-samples having very similar
clustering strengths.  Each of the conclusions drawn previously are
equally applicable to these photometrically-selected samples.

The border-defined passive samples include slightly larger numbers
  of galaxies than the more physically motivated selections presented
  earlier in this paper. The additional galaxy numbers
  enable measurements beyond $z=2$ for some of these
samples. Though highly uncertain, these measurements indicate
equivalence of clustering strength at $z>2$, as was suggested in
section \ref{disc}.

\bibliographystyle{mn2e.bst}
\bibliography{mn-jour,papers_cited_by_WH}

\begin{thebibliography}{}

\bibitem[\protect\citeauthoryear{Adelberger et 
al.}{2004}]{Adelberger04} Adelberger K.~L., Steidel C.~C., Shapley 
A.~E., Hunt M.~P., Erb D.~K., Reddy N.~A., Pettini M., 2004, ApJ, 607, 226


\bibitem[\protect\citeauthoryear{Blain et al.}{2004}]{Blain04} 
Blain A.~W., Chapman S.~C., Smail I., Ivison R., 2004, ApJ, 611, 725


\bibitem[\protect\citeauthoryear{Bond et al.}{1991}]{Bond91} 
Bond J.~R., Cole S., Efstathiou G., Kaiser N., 1991, ApJ, 379, 440 

\bibitem[\protect\citeauthoryear{Brammer et 
al.}{2009}]{Brammer09} Brammer G.~B., et al., 2009, ApJ, 706, 
L173 

\bibitem[\protect\citeauthoryear{Bruzual 
\& Charlot}{2003}]{Bruzual03} Bruzual G., Charlot S., 2003, MNRAS, 344, 1000

\bibitem[\protect\citeauthoryear{Bundy et al.}{2006}]{Bundy06} 
Bundy K., et al., 2006, ApJ, 651, 120 

\bibitem[\protect\citeauthoryear{Calzetti et 
al.}{2000}]{Calzetti00} Calzetti D., Armus L., Bohlin R.~C., 
Kinney A.~L., Koornneef J., Storchi-Bergmann T., 2000, ApJ, 533, 682

\bibitem[\protect\citeauthoryear{Carlberg et 
al.}{1997}]{Carlberg97} Carlberg R.~G., Cowie L.~L., Songaila A., 
Hu E.~M., 1997, ApJ, 484, 538 

\bibitem[\protect\citeauthoryear{Casali et 
al.}{2007}]{Casali07} Casali M., et al., 2007, A\&A, 467, 777 


\bibitem[\protect\citeauthoryear{Cimatti, Daddi, 
\& Renzini}{2006}]{Cimatti06} Cimatti A., Daddi E., Renzini A., 2006, A\&A, 453, L29

\bibitem[\protect\citeauthoryear{Cimatti et 
al.}{2008}]{Cimatti08} Cimatti A., et al., 2008, A\&A, 482, 21


\bibitem[\protect\citeauthoryear{Cirasuolo et 
al.}{2007}]{Cirasuolo07} Cirasuolo M., et al., 2007, MNRAS, 380, 
585 

\bibitem[\protect\citeauthoryear{Cirasuolo et 
al.}{2010}]{Cirasuolo10} Cirasuolo M., McLure R.~J., Dunlop J.~S., 
Almaini O., Foucaud S., Simpson C., 2010, MNRAS, 401, 1166

\bibitem[\protect\citeauthoryear{Coil et al.}{2006}]{Coil06} 
Coil A.~L., Newman J.~A., Cooper M.~C., Davis M., Faber S.~M., Koo D.~C., 
Willmer C.~N.~A., 2006, ApJ, 644, 671

\bibitem[\protect\citeauthoryear{Coil et al.}{2008}]{Coil08} 
Coil A.~L., et al., 2008, ApJ, 672, 153


\bibitem[\protect\citeauthoryear{Conselice et 
al.}{2003}]{Conselice03} Conselice C.~J., O'Neil K., Gallagher 
J.~S., Wyse R.~F.~G., 2003, ApJ, 591, 167

\bibitem[\protect\citeauthoryear{Conselice et 
al.}{2007}]{Conselice07} Conselice C.~J., et al., 2007, ApJ, 660, 
L55 


\bibitem[\protect\citeauthoryear{Conselice et 
al.}{2007}]{Conselice07} Conselice C.~J., et al., 2007, MNRAS, 381, 
962 

\bibitem[\protect\citeauthoryear{Cowie et al.}{1996}]{Cowie96} 
Cowie L.~L., Songaila A., Hu E.~M., Cohen J.~G., 1996, AJ, 112, 839


\bibitem[\protect\citeauthoryear{{Daddi}, {Cimatti}, {Renzini}, {Fontana},
  {Mignoli}, {Pozzetti}, {Tozzi} \& {Zamorani}}{{Daddi}
  et~al.}{2004}]{Daddi04}
{Daddi} E.,  {et al.},  2004, ApJ, 617, 746


\bibitem[\protect\citeauthoryear{Dressler}{1980}]{Dressler80} 
Dressler A., 1980, ApJ, 236, 351 


\bibitem[\protect\citeauthoryear{{Dye}, {Warren}, {Hambly}, {Cross}, {Hodgkin},
  {Irwin}, {Lawrence}, {Adamson}, {Almaini}, {Edge}, {Hirst}, {Jameson},
  {Lucas} \& {co-authors.}}{{Dye} et~al.}{2006}]{Dye06}
{Dye} S.,  {et al.}, 2006, MNRAS, 372, 1227

\bibitem[\protect\citeauthoryear{Einasto}{1991}]{Einasto91} 
Einasto M., 1991, MNRAS, 252, 261


\bibitem[\protect\citeauthoryear{Farrah et al.}{2006}]{Farrah06} 
Farrah D., et al., 2006, ApJ, 641, L17 


\bibitem[\protect\citeauthoryear{{Foucaud}, {Almaini}, {Smail}, {Conselice},
  {Lane}, {Edge}, {Simpson}, {Dunlop}, {McLure}, {Cirasuolo}, {Hirst}, {Watson}
  \& {Page}}{{Foucaud} et~al.}{2007}]{Foucaud07}
{Foucaud} S., {et al.},  2007, MNRAS, 376, L20

\bibitem[\protect\citeauthoryear{Foucaud et 
al.}{2010}]{Foucaud10} {Foucaud} S., {et al.}, 2010, MNRAS, in press - astro-ph/1003.2755


\bibitem[\protect\citeauthoryear{Franx et al.}{2003}]{Franx03} 
Franx M., et al., 2003, ApJ, 587, L79

\bibitem[\protect\citeauthoryear{Fujita 
\& Nagashima}{1999}]{Fujita99} Fujita Y., Nagashima M., 1999, ApJ, 516, 619

\bibitem[\protect\citeauthoryear{Furusawa et 
al.}{2008}]{Furusawa08} Furusawa H., et al., 2008, ApJS, 176, 1



 \bibitem[\protect\citeauthoryear{Gunn 
\& Gott}{1972}]{Gunn72} Gunn J.~E., Gott J.~R.~I., 1972, ApJ, 176, 1

\bibitem[\protect\citeauthoryear{Hartley et 
al.}{2008}]{Hartley08} Hartley W.~G., et al., 2008, MNRAS, 391, 
1301


\bibitem[\protect\citeauthoryear{Heinis et al.}{2007}]{Heinis07} 
Heinis S., et al., 2007, ApJS, 173, 503

\bibitem[\protect\citeauthoryear{Hewett et al.}{2006}]{Hewett06} 
Hewett P.~C., Warren S.~J., Leggett S.~K., Hodgkin S.~T., 2006, MNRAS, 367, 
454 

\bibitem[\protect\citeauthoryear{Hughes et al.}{1998}]{Hughes98} 
Hughes D.~H., et al., 1998, Natur, 394, 241

\bibitem[\protect\citeauthoryear{Ilbert et 
al.}{2006}]{Ilbert06} Ilbert O., et al., 2006, A\&A, 457, 841


\bibitem[\protect\citeauthoryear{Jing}{1998}]{Jing98} Jing 
Y.~P., 1998, ApJ, 503, L9




\bibitem[\protect\citeauthoryear{{Kong}, {Daddi}, {Arimoto}, {Renzini},
  {Broadhurst}, {Cimatti}, {Ikuta}, {Ohta}, {da Costa}, {Olsen}, {Onodera} \&
  {Tamura}}{{Kong} et~al.}{2006}]{Kong06}
{Kong} X.,  {et al.},  2006, ApJ, 638, 72

\bibitem[\protect\citeauthoryear{Landy 
\& Szalay}{1993}]{Landy93} Landy S.~D., Szalay A.~S., 1993, ApJ, 412, 64

\bibitem[\protect\citeauthoryear{{Lane}, {Almaini}, {Foucaud}, {Simpson},
  {Smail}, {McLure}, {Conselice}, {Cirasuolo}, {Page}, {Dunlop}, {Hirst},
  {Watson} \& {Sekiguchi}}{{Lane} et~al.}{2007}]{Lane07}
{Lane} K.P.,  {et al.},  2007, MNRAS, 379, L25

\bibitem[\protect\citeauthoryear{{Lawrence}, {Warren}, {Almaini}, {Edge},
  {Hambly}, {Jameson}, {Lucas}, {Casali}, {Adamson}, {Dye}, {Emerson},
  {Foucaud}, {Hewett}, {Hirst}, {Hodgkin}, {Irwin}, {Lodieu}, {McMahon},
  {Simpson}, {Smail}, {Mortlock} \& {Folger}}{{Lawrence} et~al.}{2007}]{Lawrence07}
{Lawrence} A., {et al.},  2007, MNRAS, 379, 1599

\bibitem[\protect\citeauthoryear{Le F{\`e}vre et 
al.}{2005}]{LeFevre05} Le F{\`e}vre O., et al., 2005, A\&A, 439, 845


\bibitem[\protect\citeauthoryear{Lilly 
\& Longair}{1984}]{Lilly84} Lilly S.~J., Longair M.~S., 1984, MNRAS, 211, 833

\bibitem[\protect\citeauthoryear{Lilly et al.}{1999}]{Lilly99} 
Lilly S.~J., Eales S.~A., Gear W.~K.~P., Hammer F., Le F{\`e}vre O., 
Crampton D., Bond J.~R., Dunne L., 1999, ApJ, 518, 641

\bibitem[\protect\citeauthoryear{Lilly et al.}{2007}]{Lilly07} 
Lilly S.~J., et al., 2007, ApJS, 172, 70

\bibitem[\protect\citeauthoryear{Limber}{1954}]{Limber54} Limber 
D.~N., 1954, ApJ, 119, 655 

\bibitem[\protect\citeauthoryear{Lonsdale et 
al.}{2003}]{Lonsdale03} Lonsdale C.~J., et al., 2003, PASP, 115, 
897 

\bibitem[\protect\citeauthoryear{Madau}{1995}]{Madau95} Madau 
P., 1995, ApJ, 441, 18 

\bibitem[\protect\citeauthoryear{Madgwick et 
al.}{2003}]{Madgwick03} Madgwick D.~S., et al., 2003, MNRAS, 344, 
847

\bibitem[\protect\citeauthoryear{Magliocchetti 
\& Maddox}{1999}]{Magliocchetti99} Magliocchetti M., Maddox S.~J., 1999, MNRAS, 306, 988

\bibitem[\protect\citeauthoryear{Magliocchetti et 
al.}{2008}]{Magliocchetti08} Magliocchetti M., et al., 2008, MNRAS, 
383, 1131

\bibitem[\protect\citeauthoryear{McCarthy et 
al.}{2004}]{McCarthy04} McCarthy P.~J., et al., 2004, ApJ, 614, L9

\bibitem[\protect\citeauthoryear{McCracken et 
al.}{2008}]{McCracken08} McCracken H.~J., Ilbert O., Mellier Y., Bertin E., Guzzo L., Arnouts S., Le F{\`e}vre O., Zamorani G., 2008, A\&A, 479, 321 

\bibitem[\protect\citeauthoryear{Meneux et 
al.}{2006}]{Meneux06} Meneux B., et al., 2006, A\&A, 452, 387

\bibitem[\protect\citeauthoryear{Meneux et 
al.}{2009}]{Meneux09} Meneux B., et al., 2009, A\&A, 505, 463


\bibitem[\protect\citeauthoryear{Meneux et al.}{2009}]{Meneux09} 
Meneux B., et al., 2009, arXiv, arXiv:0906.1807 

\bibitem[\protect\citeauthoryear{Mo \& White}{1996}]{Mo96} Mo H.~J., White S.~D.~M., 1996, MNRAS, 282, 347

\bibitem[\protect\citeauthoryear{Mo 
\& White}{2002}]{Mo02} Mo H.~J., White S.~D.~M., 2002, MNRAS, 336, 112

 \bibitem[\protect\citeauthoryear{Norberg et 
al.}{2002}]{Norberg02} Norberg P., et al., 2002, MNRAS, 332, 827 


\bibitem[\protect\citeauthoryear{Peebles}{1980}]{Peebles80} 
Peebles P.~J.~E., 1980, The Large-Scale Structure of the Universe. Princeton University Press, Princeton, NJ 


\bibitem[\protect\citeauthoryear{Pozzetti 
\& Mannucci}{2000}]{Pozzetti00} Pozzetti L., Mannucci F., 2000, MNRAS, 317, L17

\bibitem[\protect\citeauthoryear{Press 
\& Schechter}{1974}]{Press74} Press W.~H., Schechter P., 1974, ApJ, 187, 425



\bibitem[\protect\citeauthoryear{Roche 
\& Eales}{1999}]{Roche99} Roche N., Eales S.~A., 1999, MNRAS, 307, 703

\bibitem[\protect\citeauthoryear{{Roche}, {Almaini}, {Dunlop}, {Ivison} \&
  {Willott}}{{Roche} et~al.}{2002}]{Roche02} {Roche} N.~D., {et al.}, 2002, MNRAS, 337, 1282

\bibitem[\protect\citeauthoryear{Roche, Dunlop, 
\& Almaini}{2003}]{Roche03} Roche N.~D., Dunlop J., Almaini O., 2003, MNRAS, 346, 803



\bibitem[\protect\citeauthoryear{{Sekiguchi}, {Akiyama}, {Furusawa}, {Simpson},
  {Takata}, {Ueda}, {Watson} \& {The Sxds Team}}{{Sekiguchi}
  et~al.}{2005}]{Sekiguchi05}
{Sekiguchi} K., {et al.}, 2005, in {Renzini} A.,
  {Bender} R.,  eds, Multiwavelength Mapping of Galaxy Formation and Evolution
  {Multiwavelength Observations of the Subaru/XMM-Newton Deep Field}.


\bibitem[\protect\citeauthoryear{Sheth, Mo, 
\& Tormen}{2001}]{Sheth01} Sheth R.~K., Mo H.~J., Tormen G., 2001, MNRAS, 323, 1



\bibitem[\protect\citeauthoryear{Swinbank et 
al.}{2006}]{Swinbank06} Swinbank A.~M., Chapman S.~C., Smail I., 
Lindner C., Borys C., Blain A.~W., Ivison R.~J., Lewis G.~F., 2006, MNRAS, 
371, 465


\bibitem[\protect\citeauthoryear{Toft et al.}{2007}]{Toft07} 
Toft S., et al., 2007, ApJ, 671, 285



\bibitem[\protect\citeauthoryear{Visvanathan 
\& Sandage}{1977}]{Visvanathan77} Visvanathan N., Sandage A., 1977, ApJ, 216, 214

\bibitem[\protect\citeauthoryear{Warren et al.}{2007}]{Warren07} 
Warren S.~J., et al., 2007, MNRAS, 375, 213 

\bibitem[\protect\citeauthoryear{Wei{\ss} et 
al.}{2009}]{Weiss09} Wei{\ss} A., et al., 2009, ApJ, 707, 1201

\bibitem[\protect\citeauthoryear{Williams et 
al.}{2009}]{Williams09} Williams R.~J., Quadri R.~F., Franx M., 
van Dokkum P., Labb{\'e} I., 2009, ApJ, 691, 1879

\bibitem[\protect\citeauthoryear{Wolf, Gray, 
\& Meisenheimer}{2005}]{Wolf05} Wolf C., Gray M.~E., Meisenheimer K., 2005, A\&A, 443, 435

\bibitem[\protect\citeauthoryear{Yamada et al.}{2005}]{Yamada05} 
Yamada T., et al., 2005, ApJ, 634, 861

\bibitem[\protect\citeauthoryear{York et al.}{2000}]{York00} 
York D.~G., et al., 2000, AJ, 120, 1579

\bibitem[\protect\citeauthoryear{Zehavi et al.}{2005}]{Zehavi05} 
Zehavi I., et al., 2005, ApJ, 630, 1

\bibitem[\protect\citeauthoryear{Zirm et al.}{2007}]{Zirm07} 
Zirm A.~W., et al., 2007, ApJ, 656, 66

\bibitem[\protect\citeauthoryear{Zucca et 
al.}{2006}]{Zucca06} Zucca E., et al., 2006, A\&A, 455, 879

\end{thebibliography}

\label{lastpage}

\end{document}